\documentclass[a4paper,11pt]{article}

\usepackage[english]{babel}
\usepackage{url}
\usepackage[toc,page]{appendix}
\usepackage[squaren,Gray]{SIunits}
\usepackage{verbatim}
\usepackage{amsmath, amsthm,amsopn,amsfonts,amssymb,dsfont, esint, oldgerm,bm}
\usepackage{enumerate}
\usepackage{framed}
\usepackage{fullpage}
\usepackage{array}
\usepackage{graphicx}
\usepackage[squaren,Gray]{SIunits}

\begin{document}

\title{Thermal concentrator homogenized with solar-shaped mantle}
\author{David Petiteau$^1$, Sebastien Guenneau$^1$, Michel Bellieud$^2$, Myriam Zerrad$^1$ \\and Claude Amra$^1$}
\maketitle
\begin{center}
$^1$Aix Marseille Universit\'{e}, CNRS, Centrale Marseille, Institut Fresnel, UMR 7249, 13013 Marseille, France \\
$^2$LMGC, UMR-CNRS 5508, Universit\'{e} Montpellier II, 34095 Montpellier Cedex 5, France \\
david.petiteau@fresnel.fr
\end{center}

\begin{abstract}
We propose solar-shaped thermal concentrators designed with orthoradial layers and obtained in practice through the homogenization of an ideal thermal concentrator. Considering the spectral regime of the heat equation, we quantitatively evaluate at different pulsations the effectiveness of the homogenized concentrators by comparing the thermal flux existing in an ideal concentrator and the thermal flux in an homogenized concentrator. Dependence on the pulsation is shown to be negligible and plotting the effectiveness of the homogenized concentrators as a function of the number of orthoradial layers $N$, we determine the number of layers needed to achieve a certain effectiveness. Significantly high numbers $N$(ranging from a hundred to tens of thousands layers) are found highlighting the fact that achieving high effectiveness demands a high level of engineering of the homogenized concentrator.
\end{abstract}
\maketitle

\section{Introduction}
Following the proposal of Smith and Pendry's group to design electromagnetic cloaks and concentrators via spatially varying anisotropic media deduced from a geometric transform \cite{rahm08}, it has been shown that thermal cloaks and concentrators could be envisaged using the same techniques \cite{guenneau12}. Following these results, recent studies have been conducted to prove experimentally the feasibility of a thermal cloak \cite{schittny13, leonhardt13}. Additional studies on the control of heat diffusion have also shown that considering 
crossover from thermal diffusive effects to phonon scattering, one can design devices for heat management such as thermocrystals \cite{maldovan13} and superlattices \cite{ravichandran14}.

Regarding transformation optics applied to thermodynamics, an important issue is to engineer metamaterials so as to approach the best we can the required material parameters. A homogenization path toward multi-layered cylindrical thermal cloaks with high effectiveness has been detailed in recent work \cite{petiteau14}. In the present letter, we would like to present a similar homogenization path toward thermal concentrators with solar-shaped mantle i.e. with a structure periodic in the azimuthal direction \cite{sadeghi13}. We stress that while advances have been made in the fabrication and characterization of layered thermal cloak \cite{han13, han14}, one cannot achieve a thermal concentrator with an alternation of concentric layers, as we shall see in the sequel. On the other hand, the solar-shaped mantle structure that we here discuss provides an exact solution to approach the ideal anisotropic concentrator. Ideal and homogenized concentrators are studied in the spectral regime and it is showed that the performances of the homogenized concentrators do not depend on the pulsation. Furthermore, numerical results allow us to extrapolate on the number of layers $N$ that one should implement to achieve a certain effectiveness.
\section{Space transformation}

We consider the radii $R_1$, $R_2$ and $R_3$ such that $R_1 < R_2 < R_3$ and the two-dimensional transformation $(r', \theta') = (f(r), \theta)$ with $f$ defined as:

\begin{equation}
\label{space_transform}
\left\{\begin{array}{l l}
f(r) = \frac{R_1}{R_2}r, &\quad\text{for}\;r\in [0, R_2],\\
f(r) = \alpha r + \beta, &\quad\text{for}\;r\in [R_2, R_3],
\end{array}\right.,
\end{equation}
with $\alpha = \frac{R_3-R_1}{R_3-R_2} > 0$ and $\beta = R_3\frac{R_1-R_2}{R_3-R_2} < 0$. This transformation maps the region $0 \leq r \leq R_2$ onto $0 \leq r' \leq R_1$ (compression of thermal space) and the region $R_2 \leq r \leq R_3$ onto $R_1 \leq r' \leq R_3$ (extension of thermal space).

\begin{figure}[ht!]
\includegraphics[width=\textwidth]{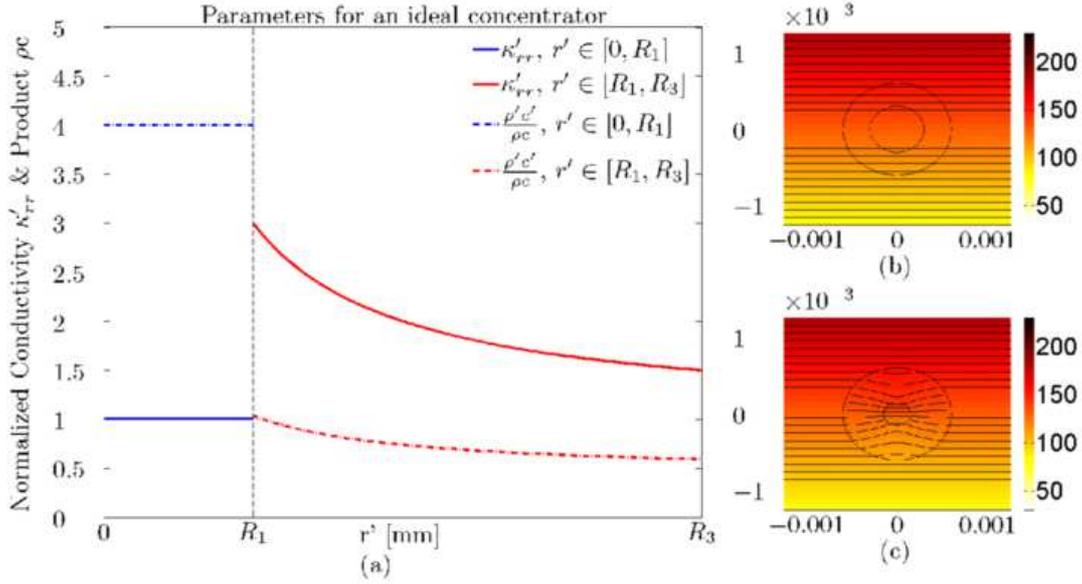}
\caption{\label{ideal_concentrator_parameters} (a) Normalized conductivity $\kappa'_{rr}$ (solid lines) and normalized product $\rho c$ parameters (dashed lines) used for an ideal concentrator. The inner boundary of the concentrator is $R_1 = 0.15$ mm and the outer boundary is $R_3 = 4R_1 = 0.6$ mm while we have $R_2 = 2R_1 = 0.3$ mm. Notice that the behaviour of $\kappa'_{\theta\theta}$ is obtained by simply calculating $\kappa'_{\theta\theta}=\frac{1}{\kappa'_{rr}}$. (b) Map of the temperature in a homogeneous medium and (c) with a concentrator, both at $\omega = 0$ rad.s$^{-1}$.}
\end{figure}

The heat equation resulting from this transformation can be written
\begin{equation}
{\rm div}\left(\underline{\kappa}'\nabla T\right) = \rho'c'\frac{\partial T}{\partial t},
\end{equation}
with $T$ the temperature and the transformed conductivity $\underline{\kappa}'$ and transformed product of heat capacity and density $\rho'c'$ written as
\begin{subequations}
\begin{align}
\underline{\kappa}' &= \kappa{\bf R(\theta)}\begin{pmatrix}
\kappa'_{rr} & 0 \\ 
0 & \kappa'_{\theta\theta}
\end{pmatrix} {\bf R(\theta)}^T, \\
\rho'c'(r) &= \rho c\frac{r}{f(r)\frac{df}{dr}(r)},
\end{align}
\end{subequations}
where ${\bf R(\theta)}$ is the rotation matrix of angle $\theta$ and we have
\begin{equation}
\kappa'_{rr} = \frac{r}{f(r)}\frac{df}{dr}(r), \; \kappa'_{\theta\theta} = \frac{1}{\kappa'_{rr}}.
\end{equation}
Writing $r' = f(r)$, we can derive $\kappa'_{rr}$ and $\kappa'_{\theta\theta}$ as a function of either $r$ or $r'$. Considering that $\kappa'_{\theta\theta} = \frac{1}{\kappa'_{rr}}$, we have
\begin{equation}
\left\{\begin{alignedat}{3}
\kappa'_{rr} &= 1 \quad& &\text{for}\left\{\begin{array}{l l}
r \in [0,R_2] \\
r'\in [0,R_1]
\end{array}\right. \\
\kappa'_{rr} &= \frac{r}{r+\frac{\beta}{\alpha}} \quad&&\text{for}\;r \in [R_2,R_3]\\
 &= \frac{r' - \beta}{r'} \quad& &\text{for}\;r' \in [R_1,R_3]
\end{alignedat}\right.,
\end{equation}
and 
\begin{equation}
\left\{\begin{alignedat}{2}
\rho'c' &= \rho c\frac{R_2^2}{R_1^2} \quad&&\text{for}\left\{\begin{array}{l l}
r \in [0,R_2] \\
r'\in [0,R_1]
\end{array}\right. \\
\rho'c' &= \rho c\frac{1}{\alpha^2}\frac{r}{r+\frac{\beta}{\alpha}} \quad&&\text{for}\;r \in [R_2,R_3]\\\\
 &= \rho c\frac{1}{\alpha^2}\frac{r'-\beta}{r'} \quad&&\text{for}\;r' \in [R_1,R_3]
\end{alignedat} \right..
\end{equation}
Introducing the Fourier transform of the temperature
\begin{equation}
u(x, \omega) = \int_{-\infty}^{\infty} T(x, t)e^{-j\omega t} \mathrm{d}t
\end{equation}
where $x$ is the space variable, $\omega$ the angular frequency and $t$ the time variable, one can write the time-harmonic heat equation as
\begin{equation}
\label{time-harmonic equation}
{\rm div}\left(\kappa\nabla u\right) +j\omega\rho c u = 0.
\end{equation}
We study this ideal concentrator behaviour for several angular frequencies $\omega$ by solving the time-harmonic heat equation (\ref{time-harmonic equation}) using the commercial finite element software COMSOL Multiphysics \textregistered. We consider a square computational domain of side length $L = 5\times 10^{-3}$ m. The medium surrounding the concentrator is made of zinc and its thermal properties are $\kappa = 121$ W.m$^{-1}$.K$^{-1}$ and $\rho c = 2.75\times 10^{6}$ J.m$^{-3}$.K$^{-1}$. The top temperature is set to $T_f = 230\degree$C and the bottom temperature is set to $T_0 = 30\degree$C. Neumann boundary conditions $\frac{\partial u}{\partial n} = 0$ (perfect insulating conditions) are set on the two other sides of our computational domain.
\begin{figure}[ht!]
\centering
	\includegraphics[width=\textwidth]{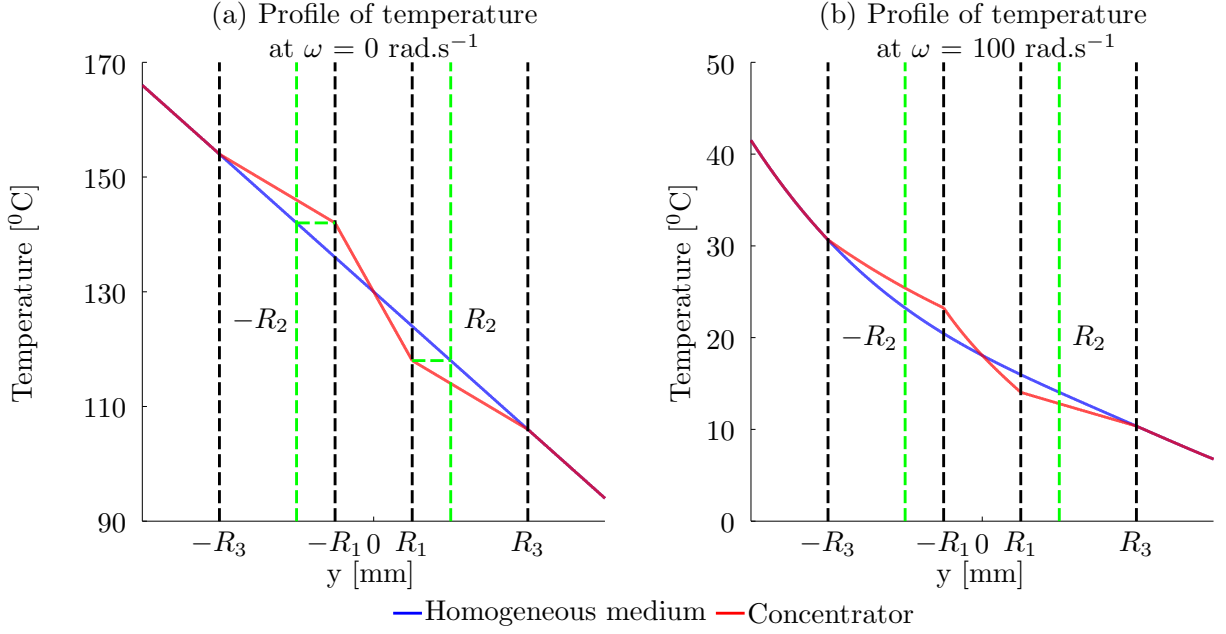}
	\caption{\label{temperature_profile} Profile of temperature along the diffusion direction ($x = 0$ and $y \in [-2.5\times 10^{-3}, 2.5\times 10^{-3}]$) at (a) $\omega = 0$ rad.s$^{-1}$ and (b) $\omega = 100$ rad.s$^{-1}$. The inner radius of the concentrator is $R_1 = 0.15$ mm while the outer radius is $R_3 = 0.6$ mm. Vertical black dashed lines denote the boundaries of the inner and outer circles. Vertical green dashed lines denote the boundaries of the initial circle of radius $R_2$ in the virtual space.}
\end{figure}

\begin{figure}[ht!]
\centering
	\includegraphics[width=\textwidth]{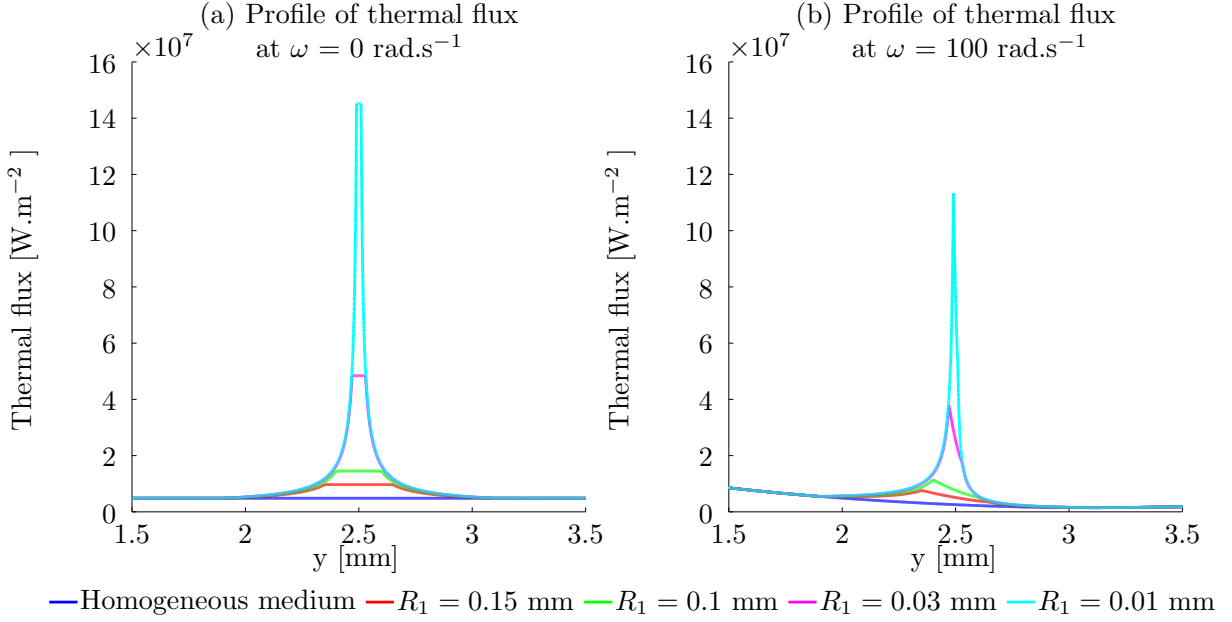}
	\caption{\label{gradient_profile_severalR1} Profile of thermal gradient along the diffusion direction ($x = 0$ and $y \in [-2.5\times 10^{-3}, 2.5\times 10^{-3}]$) with $R_1 = 0.15,\;0.1,\;0.03\;\text{and}\;0.01$ mm leading to a great enhancement of the thermal gradient in the region $0 \leq r' \leq R_1$ at (a) $\omega = 0$ rad.s$^{-1}$ and (b) $\omega = 100$ rad.s$^{-1}$.}
\end{figure}

The behaviour of $\kappa'_{rr}$ and $\rho'c'$ for $0 \leq r' \leq R_3$ is represented on Fig. \ref{ideal_concentrator_parameters}(a). One should notice that singularities do not appear in the ideal concentrator inner structure as the terms $\kappa'_{rr}$, $\kappa'_{\theta\theta}$ and $\rho'c'$ do not vanish or go to infinity in the $r'\in [R_1,R_3]$ as it would happen with an ideal invisibility cloak. Moreover, the object in the inner region of the concentrator is not chosen at will but depends on the surrounding media: its normalized radial thermal conductivity $\kappa'_{rr}$ is equal to 1 (thus the thermal conductivity of the object is $\kappa$) and its product $\rho'c'$ is proportional to $\rho c$ by a factor $\frac{R_2^2}{R_1^2}$. Both the inner region $0 \leq r' \leq R_1$ and the annulus region $R_1 \leq r' \leq R_2$ need to be filled with the according materials for the concentrator to be operational. In the end, a structure with such heterogeneous anisotropic characteristics will behave as a thermal concentrator as it will concentrate the temperature gradient that would exist on the region $0 \leq r \leq R_2$ without the concentrator, to the smaller region $0 \leq r' \leq R_1$, as it is presented in panels (b) and (c) of Fig. \ref{ideal_concentrator_parameters}. The temperature gradient is actually enhanced by a factor $\frac{R_2}{R_1}$.
The profile of the temperature along the diffusion direction at pulsations $\omega = 0$ rad.s$^{-1}$ and $\omega = 100$ rad.s$^{-1}$ is plotted in Fig. \ref{temperature_profile}. We can clearly see that the gradient of temperature existing on the segment $[-R_2, R_2]$ is constrained on a the smaller segment $[-R_1, R_1]$ at $\omega = 0$ rad.s$^{-1}$.
Fig. \ref{gradient_profile_severalR1}a and Fig. \ref{gradient_profile_severalR1}b show the variation of the thermal flux by surface unit along the diffusion direction at $\omega = 0$ rad.s$^{-1}$ and $\omega = 100$ rad.s$^{-1}$ respectively, where the thermal flux by surface unit is defined by
\begin{equation}
\label{thermal_flux_unit_surface}
\Phi = -(\underline{\kappa}\cdot\nabla T)\cdot \bf u,
\end{equation}
where $\underline{\kappa}$ is either a scalar or a matrix depending on the domain and $\bf u$ is the direction vector of heat diffusion. With $R_2 = 0.3$ mm and $R_1 = 0.15$ mm, the thermal gradient is enhanced by a factor $\frac{R_2}{R_1}$ in the region $0 \leq r' \leq R_1$.
Thus, if we choose a smaller value of $R_1$, one can achieve higher enhancement of the thermal gradient in the inner region of the concentrator as shown in Fig. \ref{gradient_profile_severalR1}a and Fig. \ref{gradient_profile_severalR1}b.

Now to obtain such a structure that mimics the thermal concentrator, we design a solar-shaped mantle in the next section.

\section{Homogenized solar-shaped mantle}
\begin{figure}[ht!]
	\centering
	\includegraphics[scale=01]{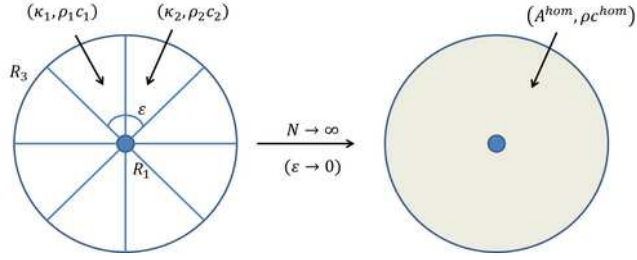}
	\caption{\label{schema_concentrateur} Schematic view of the solar-shaped homogenized concentrator. Each element is either filled with materials $(\kappa_1, \rho_1c_1)$ or $(\kappa_2, \rho_2c_2)$ and those elements are periodically laid out with $\varepsilon$ periodicity. As the number of layers $N$ goes to infinity (or equivalently, $\varepsilon \to 0$), this structure converges to an effective medium of thermal parameters $(A^{hom}, (\rho c)^{hom}$.}
\end{figure}

\begin{figure}[ht!]
	\includegraphics[width=\textwidth]{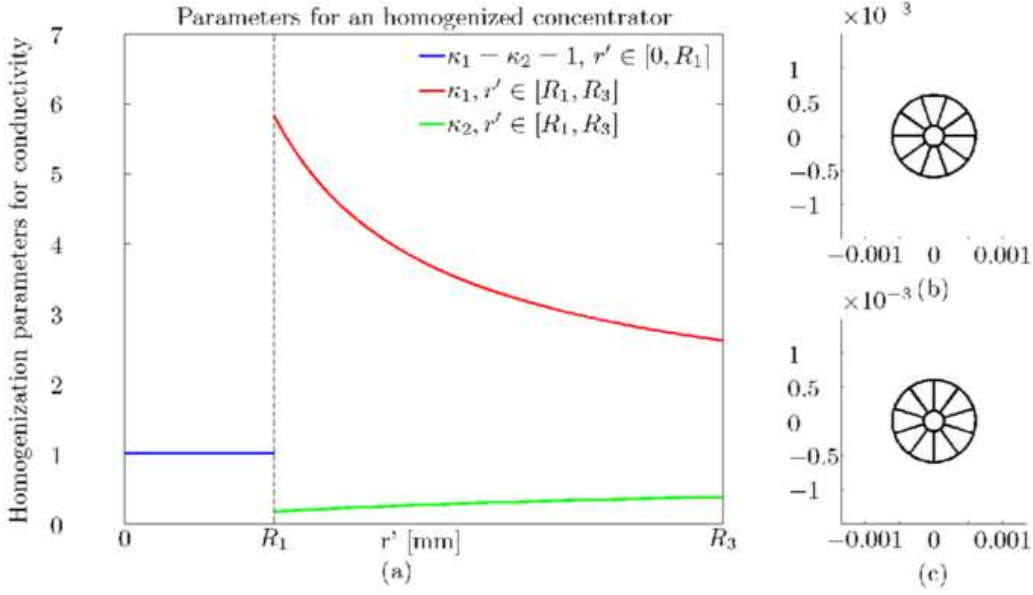}
	\caption{\label{homogenized_parameters_vs_homogenized_configuration_v2} (a) Behaviour of $\kappa_1$ and $\kappa_2$ for $r' \in [0;R_3]$. (b) Horizontal configuration of the homogenized concentrator with 10 layers: layers are implemented starting from angle $\theta = 0$. (c) Vertical configuration of the homogenized concentrator with 10 layers: layers are implemented starting from angle $\theta = \frac{\pi}{2}$.}
\end{figure}

We consider a periodic alternation of layers of respective conductivities $\kappa_1$ and $\kappa_2$ and of respective heat capacities $\rho_1c_1$ and $\rho_2c_2$ (see Fig. \ref{schema_concentrateur}) so that the overall conductivities and heat capacities can be written
\begin{subequations}
\begin{alignat}{3}
\kappa_\varepsilon &= \kappa_1(x)\mathds 1_{[0,\frac{1}{2}]}\left(\frac{\theta}{\varepsilon}\right) + \kappa_2(x)\mathds 1_{[\frac{1}{2},1]}\left(\frac{\theta}{\varepsilon}\right) \\
\rho_\varepsilon c_\varepsilon &= \rho_1c_1(x)\mathds 1_{[0,\frac{1}{2}]}\left(\frac{\theta}{\varepsilon}\right) + \rho_2c_2(x)\mathds 1_{[\frac{1}{2},1]}\left(\frac{\theta}
{\varepsilon}\right),
\end{alignat}
\end{subequations}
where $x = (x_1, x_2)$ is the position vector, $\theta = Arctan\left(\frac{x_2}{x_1}\right)$ is the angular coordinate and $\mathds 1_I$ is the indicator function of the set $I$. The function $\kappa_\varepsilon$ is periodic on the set $Y = [0,1]$ and $\varepsilon$ represents its periodicity (clearly, the thinner the layers in the cloak, the smaller the positive parameter $\varepsilon$). In this set of orthoradial layers, the time-harmonic heat equation is written
\begin{equation}
{\rm div}\left(\kappa_\varepsilon\nabla u_\varepsilon\right) +j\omega\rho c_\varepsilon u_\varepsilon = 0
\end{equation}
where $u_\varepsilon$ is the Fourier transform of the temperature and $\omega$ is the angular frequency of a periodic heat source. Following the work from G. Nguetseng and G. Allaire on two-scale convergence \cite{nguetseng89, allaire92}, it is possible to show that when $\varepsilon \rightarrow 0$, this alternation of materials behaves like an anisotropic inhomogeneous medium of conductivity
\begin{equation}
A^{hom}(x) = \begin{pmatrix}
\int_Y \kappa_0\mathrm{d}y & 0 \\
0 & \frac{1}{\int_Y\frac{1}{\kappa_0}\mathrm{d}y}
\end{pmatrix}
\end{equation}
and heat capacity
\begin{equation}
(\rho c)^{hom}(x) = \int_Y \rho_0c_0\mathrm{dy} = \frac{1}{2}(\rho_1c_1+\rho_2c_2)
\end{equation}
where $\kappa_0$ and $\rho_0c_0$ are the respective two-scale limits of $\kappa_\varepsilon$ and $\rho_\varepsilon c_\varepsilon$ when $\varepsilon \rightarrow 0$ and are written
\begin{subequations}
\begin{align}
\kappa_0 (x,y) &= \kappa_1 (x)\mathds 1_{[0,\frac{1}{2}]}(y) + \kappa_2 (x)\mathds 1_{[\frac{1}{2},1]}(y) \label{kappa_0}\\
\rho_0c_0 (x,y) &= \rho_1c_1 (x)\mathds 1_{[0,\frac{1}{2}]}(y) + \rho_2c_2 (x)\mathds 1_{[\frac{1}{2},1]}(y) \label{rhoc_0}
\end{align}
\end{subequations}
The homogenized time-harmonic heat equation is then written
\begin{equation}
\label{homogenized_heat_equation}
{\rm div}\left(A^{hom}(x)\nabla u(x)\right) +j\omega(\rho(x) c(x))^{hom}u(x) = 0
\end{equation}
where $u$ is the limit of $u_\varepsilon$ when $\varepsilon \rightarrow 0$. Then, considering (\ref{kappa_0}) and (\ref{rhoc_0}), we need to solve the system
\begin{equation}
\begin{alignedat}{1}
\label{systeme}
&\left\{\begin{array}{l l}
A^{hom} = \underline{\kappa}' \\
(\rho c)^{hom} = \rho c\det(\bf{J})
\end{array}\right. \\
\Leftrightarrow
&\left\{\begin{array}{l l}
\frac{1}{2}(\kappa_1 + \kappa_2) = \kappa'_{rr} \\ \\
\frac{2\kappa_1\kappa_2}{\kappa_1 + \kappa_2} = \kappa'_{\theta\theta} \\ \\
\frac{1}{2}(\rho c_1 + \rho c_2) = \rho c\det(\bf{J})
\end{array}\right.
\end{alignedat}
\end{equation}
for our set of materials to work as the thermal concentrator. Solving for $\kappa_1$ and $\kappa_2$ provides us with the following solution
\begin{equation}
\label{kappa1&kappa2_expression}
\left\{\begin{array}{ll}
\kappa_1 &= 1 - \frac{\beta}{r'} + \sqrt{\frac{\beta}{r'}\left(\frac{\beta}{r'}-2\right)} \\
\kappa_2 &= 1 - \frac{\beta}{r'} - \sqrt{\frac{\beta}{r'}\left(\frac{\beta}{r'}-2\right)}
\end{array}\right.
\end{equation}
The equation $(\rho c)^{hom} = \rho c\det(\bf{J})$ is not deterministic on the expression of $\rho_1c_1$ and $\rho_2c_2$. We can choose that
\begin{equation}
\label{rho1c1&rho2c2_expression}
\rho_1c_1 = \rho_2c_2 = \rho c\det({\bf J}) = \rho c\frac{1}{\alpha^2}\frac{r'-\beta}{r'}
\end{equation}
The behaviour of $\kappa_1$ and $\kappa_2$ is plotted on Fig. \ref{homogenized_parameters_vs_homogenized_configuration_v2}a.
It is interesting to notice that if one tries to obtain a thermal concentrator by homogenizing a set of concentric materials, the solutions of (\ref{systeme}) are imaginary numbers. The homogenized matrix with concentric layers is given by
\begin{equation}
A^{hom}(x) = \begin{pmatrix}
\frac{1}{\int_Y\frac{1}{\kappa_0}\mathrm{d}y} & 0 \\
0 & \int_Y \kappa_0\mathrm{d}y
\end{pmatrix}
\end{equation}
and the system we must resolve becomes
\begin{equation}
\label{systeme2}
\begin{alignedat}{1}
&\left\{\begin{array}{l l}
A^{hom} = \underline{\kappa}' \\
(\rho c)^{hom} = \rho c\det(\bf{J})
\end{array}\right. \\
\Leftrightarrow
&\left\{\begin{array}{l l}
\frac{2\kappa_1\kappa_2}{\kappa_1 + \kappa_2} = \kappa'_{rr} \\ \\
\frac{1}{2}(\kappa_1 + \kappa_2) = \kappa'_{\theta\theta} \\ \\
\frac{1}{2}(\rho c_1 + \rho c_2) = \rho c\det(\bf{J})
\end{array}\right.
\end{alignedat}
\end{equation}
This system gives the imaginary solutions
\begin{equation}
\left\{\begin{array}{ll}
\kappa_1 &= \frac{r'}{r'-\beta}\left(1+i\sqrt{\left|\frac{\beta}{r'}\left(2-\frac{\beta}{r'}\right)\right|}\right) \\
\kappa_2 &= \frac{r'}{r'-\beta}\left(1-i\sqrt{\left|\frac{\beta}{r'}\left(2-\frac{\beta}{r'}\right)\right|}\right)
\end{array}\right.
\end{equation}
Such imaginary conductivities cannot hold as physical conductivities. Thus, it is not possible to mimic a thermal concentrator with concentric layers.

\section{Numerical results}
We want to evaluate the ability of one homogenized concentrator to concentrate the thermal gradient inside the region $0 \leq r' \leq R_1$ compared to the ideal concentrator resulting from the space transformation presented in (\ref{space_transform}). We introduce the effectiveness criterion defined as the ratio between the normalized surface integral of the thermal flux in the region $0 \leq r' \leq R_1$ with the homogenized concentrator and the normalized surface integral of the thermal flux in the homogeneous medium in the region $0 \leq r \leq R_2$:
\begin{equation}
Thermal\;flux\;ratio = \frac{\frac{1}{\pi R_1^2}\iint_{R_1} \Phi(homogenized\;concentrator)\mathrm{d}S}{\frac{1}{\pi R_2^2}\iint_{R_2} \Phi(homogeneous\;medium)\mathrm{d}S}
\end{equation}
where $\Phi$ is defined in (\ref{thermal_flux_unit_surface}). As stated before, the ratio between thermal flux with the ideal concentrator in the region $0 \leq r' \leq R_1$ and the thermal flux in the homogeneous medium in the region $0 \leq r \leq R_2$ is equal to $\frac{R_2}{R_1}$. Therefore, the better the homogenized concentrator, the closer the $Thermal\;flux\;ratio$ to $\frac{R_2}{R_1}$.

\begin{figure}[ht!]
	\includegraphics[width=1.0\textwidth]{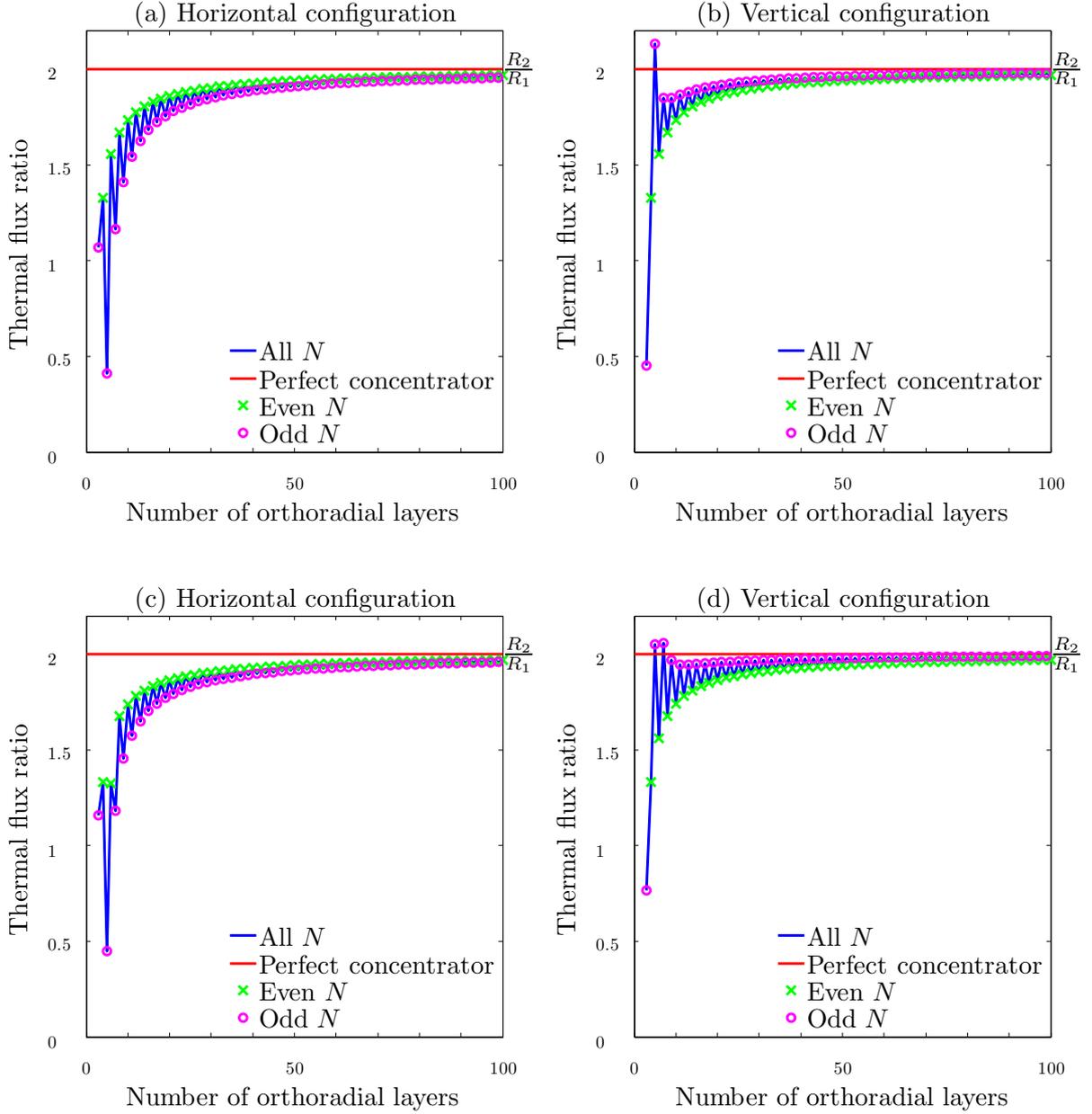}
	\caption{\label{thermal_flux_convergence_horizontal_vs_vertical_omega=0} Convergence of the thermal gradient obtained in the region $0 \leq r' \leq R_1 = 0.15$ mm with homogenized concentrators of orthoradial layers ranging from $N = 3$ to $N = 100$ at $\omega = 0$ rad.s$^{-1}$ in (a) horizontal configuration and (b) vertical configuration and at $\omega = 100$ rad.s$^{-1}$ in (c) horizontal configuration and (d) vertical configuration.}
\end{figure}

Thus we performed simulations for homogenized concentrators ranging from $N = 3$ to $N = 100$ orthoradial layers and calculated the $Thermal\;flux\;ratio$ obtained in the region $0 \leq r' \leq R_1$ at $\omega = 0$ rad.s$^{-1}$ and $\omega = 100$ rad.s$^{-1}$. We distinguished two configurations for the homogenized concentrators: a horizontal configuration and vertical configuration as represented in Fig. \ref{homogenized_parameters_vs_homogenized_configuration_v2}b and \ref{homogenized_parameters_vs_homogenized_configuration_v2}c. Horizontal configuration corresponds to the configuration where orthoradial layers are implemented from angle $\theta = 0$ whereas vertical configuration denotes orthoradial layers implemented from angle $\theta = \frac{\pi}{2}$. Depending on the symmetry of the homogenized concentrator, heat fluxes won't experience the same structure. Results are presented in Fig. \ref{thermal_flux_convergence_horizontal_vs_vertical_omega=0}.
As we can see, both horizontal and vertical configurations of the homogenized concentrators give oscillations of the thermal flux ratio. These oscillations appear both at $\omega = 0$ rad.s$^{-1}$ and $\omega = 100$ rad.s$^{-1}$ and converge towards the same limit $\frac{R_2}{R_1}$. This highlights the fact that the convergence of the $Thermal\;flux\;ratio$ is independent of the angular frequency $\omega$ (other values of $\omega$ would give the similar oscillations converging towards the same limit). This result is to be expected since the angular frequency $\omega$ does not appear in the homogenized parameters as we can see from equations (\ref{homogenized_heat_equation}), (\ref{kappa1&kappa2_expression}) and (\ref{rho1c1&rho2c2_expression}). Now, independent of $\omega$, even values of $N$ in horizontal and vertical configuration give significantly the same value of $Thermal\;flux\;ratio$ for a given number $N$ of layers whereas odd values of $N$ lead to different $Thermal\;flux\;ratio$ in horizontal and vertical configurations. This result is explained by the fact that, for even values of $N$, the homogenized concentrator is perfectly symmetric with respect to the heat diffusion direction, in horizontal as in vertical configuration. As a comparison, odd values of $N$ do not ensure the symmetry of the homogenized concentrator. These results illustrate the fact that the effectiveness of homogenized concentrators depends substantially on the layout of the orthoradial layers. Such considerations of symmetry do not appear with the homogenization of cloaks since these involve only concentric layers meaning perfect symmetry whatever the direction of the heat diffusion. We see however that the difference of oscillations between even and odd values of $N$ tend to zero as $N$ increases as one should expect.
\begin{figure}[ht!]
	\includegraphics[width=\textwidth]{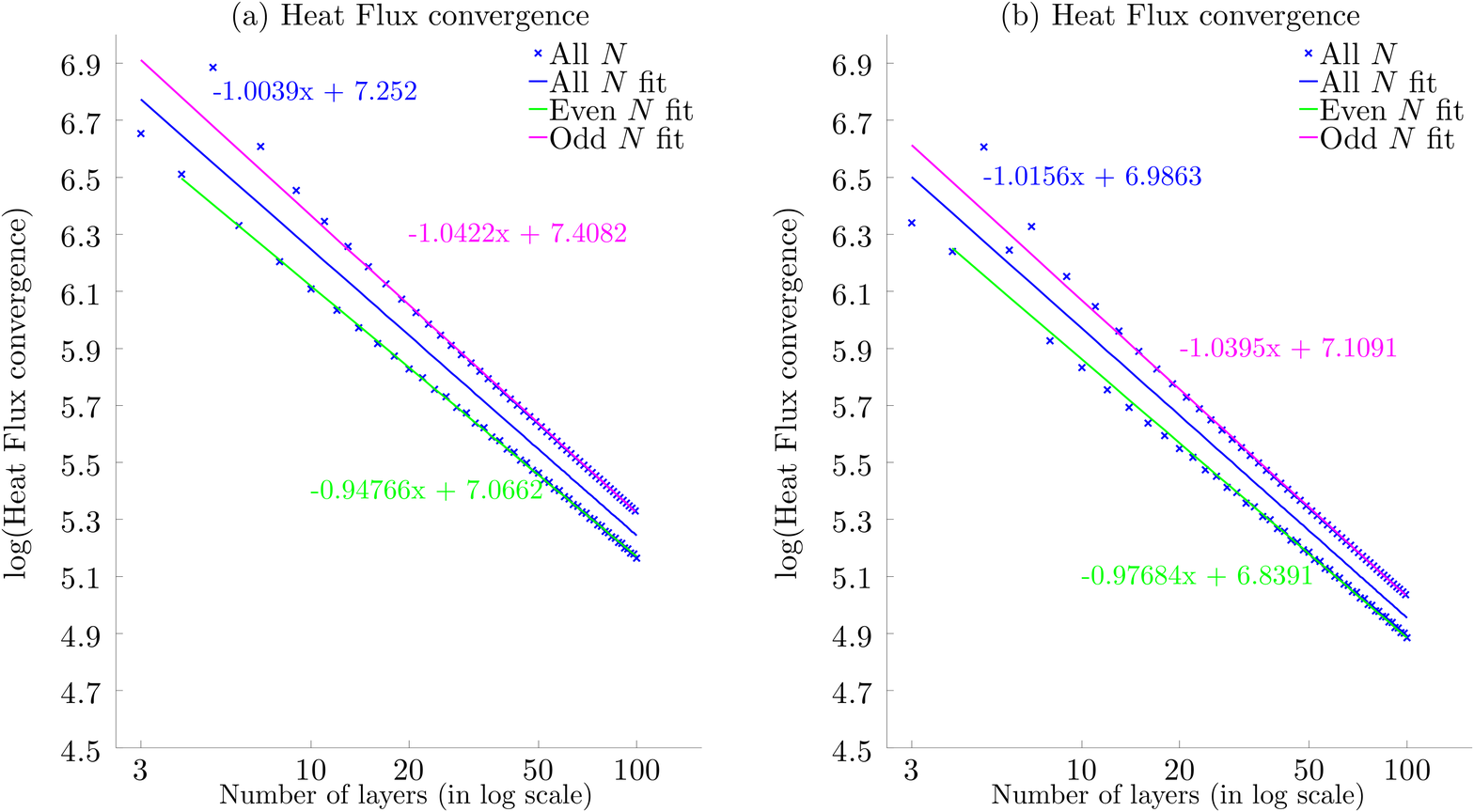}
	\caption{\label{Heat_flux_convergence_vs_N_log_scale} Heat flux convergence as a function of ${\rm log}(N)$ for the horizontal configuration at (a) $\omega = 0$ rad.s$^{-1}$ and (b) (a) $\omega = 100$ rad.s$^{-1}$. Linear fitting has been applied to all values of $N$, and even and odd values of $N$ separately.}
\end{figure} 
Now to understand further the behaviour of the homogenized concentrators, we define the quantity
\begin{equation}
\begin{alignedat}{1}
&Heat\;Flux\;convergence = \\
&\frac{1}{\pi R_1^2}\left|\iint_{R_1}\Phi(homogenized\;concentrator)\mathrm{d}S - \iint_{R_1} \Phi(perfect\;concentrator)\mathrm{d}S\right|.
\end{alignedat}
\end{equation}
This quantity is going to zero as $\varepsilon \to 0$ (i.e $N$ going to infinity). We can use the result \cite{pakhnin12}
\begin{equation}
\label{homogenization_convergence}
||u-u_\varepsilon||_{L^2} = \sqrt{\int_\Omega |u-u_\varepsilon|^2 \mathrm{dx}} \leq C(\Omega)\varepsilon
\end{equation}
where $u_\varepsilon$ denotes the temperature in the computing domain $\Omega$ for a $N$-layer concentrator and $u$ is the limit of $u_\varepsilon$ when $\varepsilon \to 0$. Plotting the decimal logarithm of the $Heat\;Flux\;convergence$ as a function of ${\rm log}(N)$, we can verify if the behaviour of $Heat\;Flux\;convergence$ as $N$ goes to infinity is consistent with (\ref{homogenization_convergence}) for both $\omega = 0$ rad.s$^{-1}$ and $\omega = 100$ rad.s$^{-1}$. Results are illustrated in Fig. \ref{Heat_flux_convergence_vs_N_log_scale}.

Linear fitting has been applied to all values of $N$, and even and odd values of $N$ separately. We find that
\begin{equation}
{\rm log}(Heat\;Flux\;convergence) \approx -{\rm log}(N) + C
\end{equation}
for the three sets of points in Fig. \ref{Heat_flux_convergence_vs_N_log_scale}a and Fig. \ref{Heat_flux_convergence_vs_N_log_scale}b, confirming the behaviour expected from (\ref{homogenization_convergence}). As seen above, the homogenization convergence is again independent of angular frequency $\omega$ as one should expect. Using the linear fit for all $N$ values (blue curve), we can extrapolate to find the required number $N$ of layers to achieve a desired value of $Heat\;Flux\;convergence$. 

\begin{figure}[ht!]
	\includegraphics[width=\textwidth]{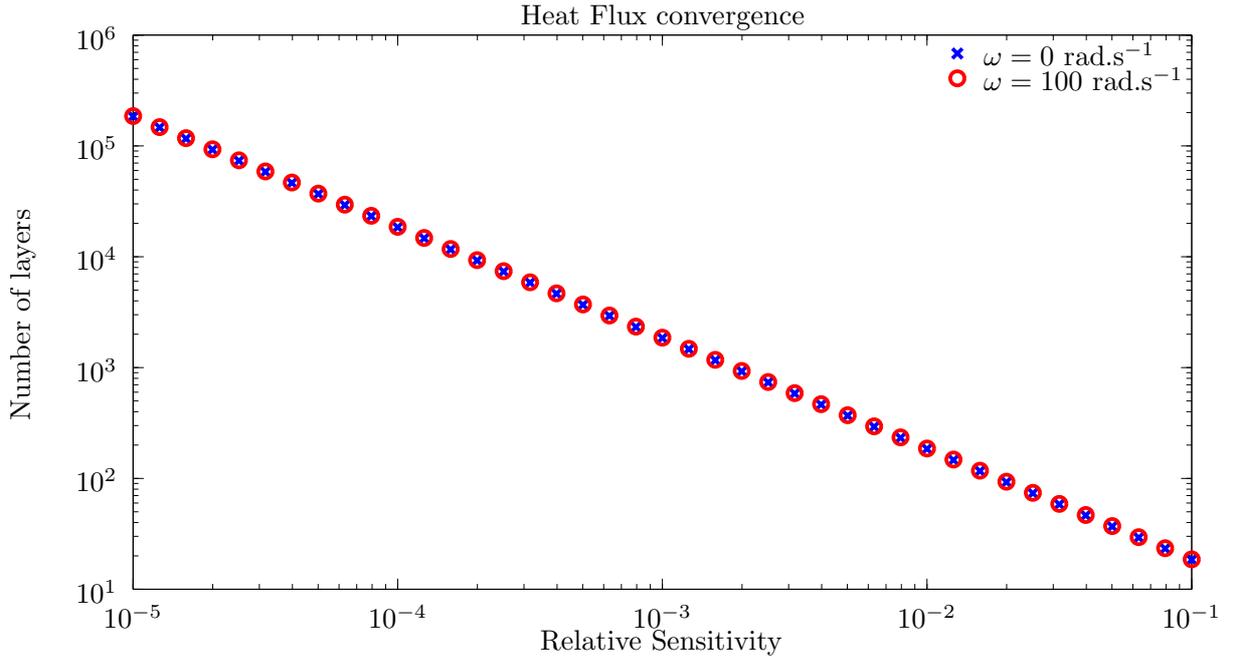}
	\caption{\label{heat_flux_convergence_extrapolation} Number of layers as a function of relative sensitivity at $\omega = 0$ rad.s$^{-1}$ (blue) and $\omega = 100$ rad.s$^{-1}$ (red) in the horizontal configuration.}
\end{figure}

The relative sensibility is defined by the ratio of the $Heat\;Flux\;convergence$ and the normalized heat flux $\frac{1}{\pi R_1^2}\iint_{R_1} \Phi(perfect\;concentrator)\mathrm{d}S$. The number $N$ of layers as a function of relative sensibility for both $\omega = 0$ rad.s$^{-1}$ and $\omega = 100$ rad.s$^{-1}$ is plotted on Fig. \ref{heat_flux_convergence_extrapolation} in the case of horizontal configuration. 
We see for instance that to reach a relative sensitivity of $10^{-3}$, one would need to implement a homogenized concentrator with nearly 2000 layers. A relative sensitivity of $10^{-4}$ would need an even higher number of layers, close to 20000. Therefore, we can determine the order of magnitude of the number of layers that is needed to reach a certain effectiveness. However, the order of magnitude of the number of layers on Fig. \ref{heat_flux_convergence_extrapolation} highlights the fact that achieving high effectiveness demands a high level of engineering of the homogenized concentrator.
\section{Conclusion}
While concentric multilayers do not allow to reach perfect heat concentrators, it was shown how the homogenization of orthoradial multilayered structures allows one to achieve an ideal anisotropic heat concentrator. The complete design of such structures was given and we calculated the temperature and flux variations within the thermal device at different pulsations. Furthermore, the quantitative effectiveness of the homogenized concentrators was numerically calculated versus the number of orthoradial layers and it was shown that the performances of the multi-layered concentrators are independent of the pulsation. Extrapolation on the numerical results allows one to adapt the complexity of the concentrator once the required performances are chosen and fairly high numbers of layers are found. This study emphasizes the fact that engineering devices for the heat management requires optimal designs but paves the way towards highly effective thermal metamaterials.

\section{Acknowledgements}
We acknowledge the support of the Direction G\'{e}n\'{e}rale de l'Armement (DGA) and funding from ANR through INPACT project.

\end{document}